\documentclass[twocolumn,showpacs,prl,amsmath,amssymb]{revtex4}
\usepackage{graphicx}
\usepackage{dcolumn}
\usepackage{bm}

\begin{document}

\title{Nearly free electrons in the layered oxide superconductor Ag$_{5}$Pb$_{2}$O$_{6}$}

\author{Mike Sutherland$^1$, P.D.A. Mann$^1$, Christoph Bergemann$^1$,  
Shingo Yonezawa$^{2}$, Yoshiteru Maeno$^{2,3}$}

\affiliation{$^1$Cavendish Laboratory, University of Cambridge, Madingley Road, Cambridge CB3 0HE, UK}
\affiliation{$^2$Department of Physics, Graduate School of Science,~Kyoto~University,~Kyoto 606-8502, Japan}
\affiliation{$^3$International Innovation Center, Kyoto University, Kyoto, 606-8501, Japan }

\begin{abstract}
 We present first measurements of quantum oscillations in the layered oxide superconductor 
 Ag$_5$Pb$_2$O$_6$.  From a detailed angular and temperature dependent study of the dHvA effect 
 we determine the electronic structure
 and demonstrate that the electron masses are very light, $m^*$ $\sim$ 1.2 $m_e$.  The Fermi 
 surface we observe is essentially that expected of nearly-free electrons---establishing 
 Ag$_5$Pb$_2$O$_6$ as the first known example of a monovalent, nearly-free electron superconductor 
 at ambient pressure.

 \end{abstract}

\maketitle

	Almost 100 years of low temperature research has established superconductivity as one of 
  nature's favoured electronic ground states.  The list of superconductors 
  is both long and diverse --- multivalent elements \cite{Anderson}, complex organic 
	molecules \cite{Williams}, heavy fermion metals \cite{Joynt} and layered 
  oxides \cite{Cava} are just a few examples of materials in which electrons form a coherent state 
  of Cooper pairs upon cooling. In many of these systems theoretical understanding can be
  hindered by electronic complexity: even elements 
  like Al or Pb, though free electron like, have a number of bands which cross the Fermi 
  energy $E_F$, resulting in intricate Fermi surfaces. 
  
  It is perhaps the desire to study superconductivity in the simplest environment
  possible 
  that has driven a long history of exploration in the most primitive of all metallic systems: 
  the  `jellium-like'  alkalis \cite{Lang,Juntunen,Jansen2} and the noble metals \cite{Ono,Buhrman}. 
  Despite 
  significant effort, measurements on samples of exceptionally
  high purity at temperatures in the micro-Kelvin range have yielded no signs of 
  superconductivity.  
  It appeared until now, empirically at least, that nature does not provide us with an example of a single 
  band, monovalent, nearly-free electron material that superconducts at ambient pressure 
  \cite{footnote}.
  
    In this Letter we end this search by reporting the observation of quantum 
    oscillations in magnetization (the de~Haas-van Alphen or dHvA effect \cite{Shoenberg}) in the 
    layered oxide superconductor Ag$_5$Pb$_2$O$_6$. Our measurements demonstrate for the first time 
    a material 
    that possesses a nearly-free electron Fermi surface, $and$ superconductivity at low temperatures.  
    In doing so we also resolve a decade old controversy regarding the electronic 
    configuration of this compound, and provide insight into the origins of the 
    rather unusual range of observed $T^2$ resistivity.

  Ag$_5$Pb$_2$O$_6$ was first grown in the 1950s by Bystr\"om and Evers 
  \cite{Bystrom}, but it is only in the past few years that single crystal samples have been 
  synthesized \cite{Abe,Yonezawa}.  The structure of this material is an interesting one --- 
  consisting of planar Kagom\'{e} lattices of 
  silver atoms separated by PbO$_6$ octahedra, and threaded with chains of silver atoms running along 
  the $c$-axis \cite{Jansen1}.  Some of us have recently 
  reported bulk superconductivity, with a $T_c$ of 52mK \cite{Yonezawa2}, establishing Ag$_5$Pb$_2$O$_6$
  as the first example of a layered silver oxide superconductor. 
  
    The renewed interest in this system has brought to light a number of outstanding issues.  
    Transport measurements have revealed 
    that the material is a good metal, with a residual resistivity as low as 1.5 $\mu\Omega$cm and 
    9.7 $\mu\Omega$cm for transport within and perpendicular
      to the planes, respectively.  What is very surprising is that the $T^2$ regime of the 
      resistivity, usually associated with the Fermi liquid state at low temperatures, persists 
      all the way up to room temperature.  The pronounced $T^2$ term leads to a Kadowaki-Woods (KW) 
      ratio that is 
      quite high, about 14 times the usual value of 1.0 $\times$ 10$^{-5} \mu\Omega$cm (K mol/mJ)$^2$
      expected in correlated systems \cite{Kadowaki}.  This is particularly unusual as it has been
      suggested that neither 
      strong electron-electron correlations nor coupling to high frequency optical phonon modes is
      expected from measurements of the specific heat \cite{Yonezawa}.  
      
  Secondly, the origin of metallic transport itself and the electronic structure of the system 
  have proven to be 
  controversial.  Jansen $et. al.$ \cite{Jansen1} and Brennan and Burdett 
  \cite{Brennan} have proposed conflicting models of the valence state formulation and 
  electron distribution in Ag$_5$Pb$_2$O$_6$, and band structure calculations 
  \cite{Brennan, Shein, Oguchi} yield strikingly different Fermi surfaces.
  
  \begin{figure} \centering
  \resizebox{8.5cm}{!}{
  \includegraphics[angle=0]{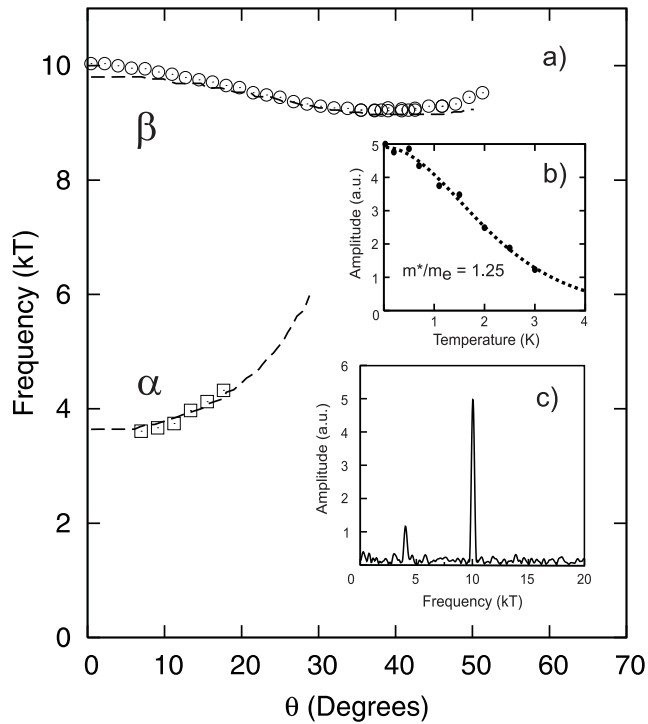}}
  \caption{\label{fig:res}  a)  Measured dHvA frequencies in Ag$_5$Pb$_2$O$_6$ as the angle between the 
  magnetic field and $c$-axis is varied.  The data for the rotation study was taken at T $\sim$ 20 
  mK and shows two branches labelled $\alpha$ and $\beta$.
  The dashed lines represent dHvA frequencies based on the band structure calculations
  of Oguchi \cite{Oguchi} which predict nearly-free electron behaviour.
  b) The dependence of the amplitude of the $\beta$ branch at $\theta$ = 0$^{\circ}$ on temperature.  The 
  line through the data is a fit to the standard Lifshitz-Kosevich formula, yielding $m^*$=1.25 $\pm$ 
  0.10 $m_e$.  The $\alpha$ branch gives $m^*$=1.1 $\pm$ 
  0.2 $m_e$ for the same orientation (not shown).
  c) A sample dHvA spectra taken at $T \sim 20$ mK and $\theta$ $\sim$ 20$^{\circ}$.}
  
\end{figure}

  To address these issues we used measurements of dHvA oscillations to firmly establish the 
  electronic structure of this material.  Our high quality single crystals were grown in Kyoto 
  \cite{Yonezawa} using an AgNO$_3$ self-flux method with a 5:1 ratio of Ag to Pb.
  Our samples were rod 
  shaped, approximately 1 mm in length and 50 $\mu$m in diameter.  The experiments were carried 
  out in a low-noise superconducting magnet system using field sweeps between 15 -- 18 T, at temperatures ranging
  from 20 mK to 3 K. Detection of the dHvA signal was via a 
  field modulated AC technique.  A 6 Hz modulation field of 11 mT in strength was applied and the 
  second harmonic of the voltage of the pickup coils was recorded, essentially providing a measure
  of $\partial^2M/\partial B^2$.  Two primary dHvA frequencies were observed, and their  
  dependence on the angle between the applied field and the $c$-axis (the polar angle, $\theta$) 
  is shown in Fig. 1.  The alignment of the sample allows 
  for at most a few degrees error in the angular position.  
  
  What is immediately obvious from the data is that the high frequency ($\beta$) branch is very weakly 
  dependent on $\theta$, indicating extremal electron orbits which do not vary significantly upon 
  rotation. 
  When the field was aligned close to the $c$-axis ($\theta =0$) we observe a second, lower frequency
  ($\alpha$) branch, that rises rapidly as $\theta$ was increased.  No oscillations were observed 
  for $\theta \gtrsim 52^\circ$.
  
  The data in Fig.~1 provide a straightforward test of
  band structure calculations.   Brennan and Burdett \cite{Brennan} have
   used the extended H\"uckel implementation of tight-binding theory to compute the electronic
   structure, and the rather complex Fermi surface they arrive at is reproduced in Fig.~$2c$.  
   In their calculations, strong antibonding interactions between silver and the lead oxide octhedra 
   cause the main cylindrical body of the FS to splay, with multiple arms radiating along the central 
   plain and extending all the 
   way to the Brillouin 
   zone edges.  The predicted dHvA signal would then include contributions from the large electron 
   orbits about the main body of the cylinder, with F $\sim$ 7 kT on-axis, as well as 
   smaller hole orbits between the arms with F $\sim$ 1.5 kT on-axis.  
         
  Our observations clearly contradict this picture.  The frequency of the large
   orbit is some 40 \% greater than predicted, and its angular dependence does not vary as 
   1~/cos $\theta$ as expected for a cylinder.  The smaller frequency is too large to be 
   attributed to the `armpit' hole orbits, and there is no obvious way to account for an orbit of 
   this size.
   It is apparent then that this initial attempt at band structure calculations completely fails
   to describe the actual electronic structure of Ag$_5$Pb$_2$O$_6$, which is rather unusual given the 
   accuracy such calculations typically have.
       
    The relative simplicity of our results allows us to attempt to reconstruct a plausible FS
    directly from the data. In the simplest approach, we can assume that the electrons are predominantly of 
    nearly-free $s$ 
    character. Given the dimensions of the Brillouin zone and assuming half 
 filling, 
  the resulting FS is spherical, with radius 0.533 \AA$^{-1}$.  Such a sphere would 
  yield an dHvA frequency of 9.34 kT, which already describes our data rather well.  The $\beta$ 
  branch varies by only $\pm$ 4 \% upon rotation between 0 to 60 $^{\circ}$, 
  with an average value of 9.52 kT.
  
  The fact that the radius of this sphere is greater than the distance from the midpoint of the Brillouin
  zone to the $c^*$ axis face implies that there exists a neck orbit at 
  the top and bottom of the zone, in analogy to the cases of elemental Cu, Au or Ag. Identifying this orbit
  with the $\alpha$ branch in Fig. 1 we estimate the neck radius to be 0.322 \AA.  In combining
  these two results, the picture that emerges is a Fermi surface that is exceedingly simple: an oblate
  spheroid with a neck extending along the $c^*$-direction. 
 
 This rudimentary scenario is supported by more recent band structure calculations by Oguchi 
 \cite{Oguchi}, 
 which predict a half filled conduction
  band composed predominantly of Pb-6$s$ and O-2$p$ orbitals with a single nearly-free electron per
  formula unit.
  The calculated dHvA 
  frequencies in this model are shown by the dashed line in Fig.1, in remarkable agreement with our 
  data.  The 
  convergence of calculation and experimental data is solid evidence that the electronic structure 
  of Ag$_5$Pb$_2$O$_6$ is indeed nearly-free electron like.
  
         \begin{figure} \centering
         \resizebox{7.8cm}{!}{
         \includegraphics[angle=0]{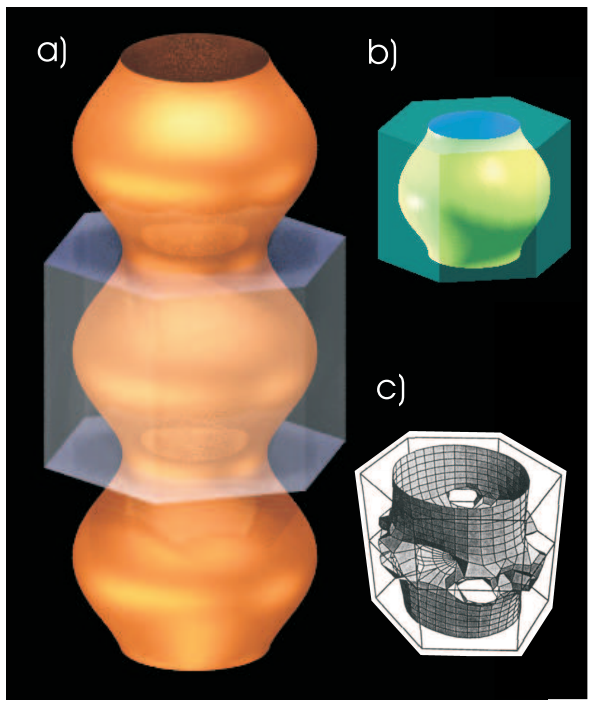}}
         \vspace{.0cm}
         \caption{\label{fig:res}(colour) a) The FS of Ag$_5$Pb$_2$O$_6$
         as measured by dHvA, shown in the periodic zone scheme, with the translucent box
         corresponding to the first Brillouin zone of the hexagonal lattice. b) The FS
         predicted by the band structure calculations of Oguchi \cite{Oguchi}. 
         c) The proposed FS of Brennan and Burdett \cite{Brennan}. }
         
     \end{figure}
 
We may now set about parameterizing the shape of the FS in full detail based on our experimental data.  
At an arbitrary angle $\theta$, the cross sectional belly area sampled by dHvA is approximately
elliptical with area $\pi k_{\rm minor}k_{\rm major}$ where $k_{\rm minor}$ and $k_{\rm major}$ correspond to the minor
and major axes of an ellipse in reciprocal space.  When $\theta$ = 0$^{\circ}$, $k_{\rm minor}$= $k_{\rm major}$
= 0.552 \AA, and the cross section is a circle.  As the sample is rotated, $k_{\rm major}$ increases 
while $k_{\rm minor}$ remains unchanged, so for a rotation angle $\theta$ we can uniquely determine 
$k_{\rm major}$ from the area of the cyclotron orbit $A_c$. $k_{\rm major}$ may be written in terms of its in
plane ($k_x$) and out of plane ($k_z$) components, which we define by $k_x=A_{c}$cos$(\theta)/\pi k_{\rm minor}$
and $k_z=A_{c}$sin$(\theta)/\pi k_{\rm minor}$.
 
Applying this analysis to both the belly and neck orbits we can then find the dependence of the 
radius
of the FS ($k_x$) as a function of the distance along the c$^*$ axis ($k_z$) which may be suitably 
parameterized in cylindrical harmonics \cite{Bergemann}.  To a high
degree of accuracy, we find that $k_x = 0.444 + 0.112\,\cos(k_z c) - 0.005\,\cos(2 k_z c)$ 
 \AA$^{-1}$, which we use to construct the full three 
dimensional FS of Ag$_5$Pb$_2$O$_6$ in Fig. 2a. The resemblance to the FS derived from the recent band 
structure calculations of Oguchi \cite{Oguchi}, shown in Fig. 2b is impressive, confirming
the electronic structure of this material.
       
   We may gain further confidence in the validity of this picture by considering the dependence of 
   the experimental signal {\em amplitude\/} on angle in Fig.\ 3.
   The higher frequency oscillations are dramatically enhanced when the field is 
   oriented at angles close to 43$^\circ$ from the $c$-axis.  This effect routinely arises in 
   quasi-2D systems possessing cylindrical Fermi surfaces with a periodic warping along
   the $c$-axis \cite{Yamaji}.  For a system with a $c$-axis lattice constant $c$, and an average 
   cylinder diameter
   $k_F$, such an enhancement is expected  
   at the Yamaji angle, given by $\theta_Y = \arctan(\xi ck_F)$ where
   $\xi \simeq 2.405$. At this angle, all of the cyclotron 
   orbits about the FS have identical areas, causing a large increase in the signal amplitude.  In the 
   present case, the deviation from an ideal cylinder is clearly large, but a remnant Yamaji effect
   is still expected.  
   Using an average 
   $k_F$ = 0.437 \AA \hspace{2pt}estimated from the maximum 
   (belly orbit) and minimum (neck orbit) dHvA frequencies, $\theta_Y=40^\circ$, in good agreement with 
   the peak position in Fig.\ 3.  Additionally, a beating frequency was observed as this angle was 
   approached, which disappeared completely at $\theta_Y$ as anticipated in the Yamaji picture.  Such
  observations serves as further confirmation of the FS depicted in Fig.~2a.
  
  We have also studied the 
  temperature dependence of the quantum oscillations and find that the effective electron masses are 
  very light.  With B $\parallel$ $c$ the standard 
  Lifshitz-Kosevich equation \cite{Lifshitz} yields $m^*$ = 1.25 $\pm$ 0.10 $m_e$ for the $\beta$ 
  branch
  (shown in the inset of Fig.~1) and $m^*$ = 1.1 $\pm$ 0.2 $m_e$ for the $\alpha$ branch.
  Masses of this magnitude effectively rule out any sizeable electron-electron correlation 
  effects in this system.  This number tallies quite well with the observed low-T specific heat,
  $C_{\rm el}/T = 3.4\ $ mJ/mol K$^2$ \cite{Yonezawa}.  Using the free electron approximation above, 
  we 
  estimate
  $C_{\rm el}/T = \pi^2 m^* k_B^2 N_A / \hbar^2 k_F^2 = 3.85$ mJ/mol K$^2$ where we have taken 
  $k_F = 0.53$ \AA\ to account for the slightly truncated spherical shape of 
  the Fermi surface.  This good agreement implies that there are no additional 
  undetected sheets on the FS.
 
                \begin{figure} \centering
                \resizebox{7.25cm}{!}{
                \includegraphics[angle=0]{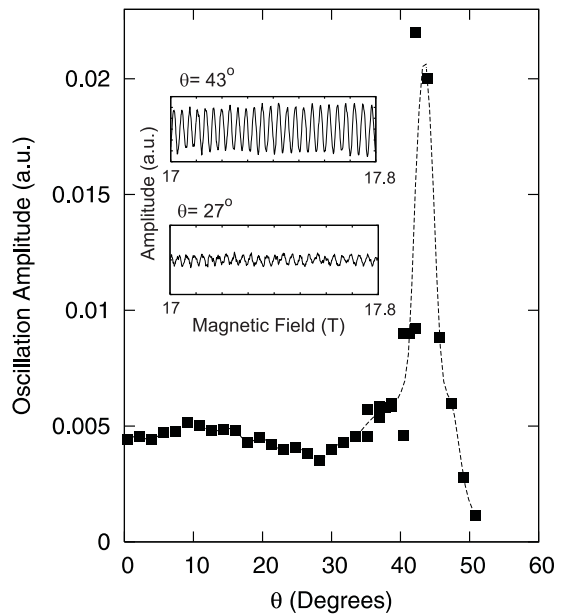}}
                \caption{Main panel:  The evolution of the amplitude of dHvA oscillations
                as a function of polar angle $\theta$.  The vertical axis is the Fourier amplitude
                of the large frequency oscillation, in arbitrary units.  
                Inset:  Examples of dHvA field sweeps on Ag$_5$Pb$_2$O$_6$, taken for angles 
                $\theta$ = 27$^\circ$ and 43$^\circ$.  The vertical axis is the pickup signal (in arbitrary units, 
                same scale for both sweeps) at the second harmonic of the excitation frequency. 
                A large enhancement in the oscillation amplitude is seen at the Yamaji angle.}
                
       \end{figure}
       
  The very low effective electron mass also implies a small electron-phonon coupling constant.
  In a simple model, the effect of phonons on electron energy levels within $\hbar 
  \omega_D$ of $E_F$ is to modify the density of states by a factor 1+$\lambda$ . This means
  that in systems with strong electron-phonon coupling the specific heat should deviate considerably
  from the nearly-free electron expectation.  The band structure calculations of Oguchi \cite{Oguchi} 
  predict a band mass $m_B$ of 1.2 $m_e$ for the $\beta$ branch at $\theta=0^\circ$, which is identical 
  within error to our measured value of $m^*$.  This implies that $\lambda$ is exceedingly small, 
  {\em at most\/} 0.1, and can 
  thus be taken as strong evidence that the phonons do not play a significant role in 
  renormalizing electronic properties. 
  
    The small effective mass also makes the observation of superconductivity in this 
    system rather remarkable.  The value of $\lambda$ in Ag$_5$Pb$_2$O$_6$ is in fact 
    lower than that observed in other nearly-free electron metals.  Copper for instance
    has an $m^*$ = 1.3 $m_e$ with $\lambda$ = 0.3 for some orbits (\cite{Shoenberg} and references 
    therein).  This
    is puzzling---if the electron-phonon
    coupling is so weak in this system, why should Ag$_5$Pb$_2$O$_6$ be a superconductor when the 
    noble metals and the alkalis are not?  
    
    One intriguing possibility is that the electron-phonon
    coupling in this system is highly anisotropic, varying significantly in strength as one moves around
    the Fermi surface.  There is evidence that this may indeed be the case from our measurements of the 
    effective mass.  The neck orbits are predited to have an $m_B$ = 0.68 $m_e$ when B $\parallel$ 
    $c$ \cite{Oguchi}, yet the measured mass from dHvA is 1.1 $\pm$ 0.2 $m_e$ leading to
    a $\lambda$ of between 0.3 and 0.9.  This is much larger than the $\lambda$ for the belly orbit,
    suggesting that perhaps superconductivity arises predominantly from electrons near the necks
    of the Fermi surface shown in Fig.\ 2.

  These considerations only serve to deepen the mystery of the broad range of $T^2$ resistivity in 
  Ag$_5$Pb$_2$O$_6$.  The rather large value of the KW ratio suggests that the $T^2$ term does not
  arise from electron-electron correlations in the usual manner, a fact confirmed by our direct
  measurement of the effective mass.  Similarly, the absence of large electron-phonon coupling inferred 
  from our data also rules out a the scenario of Gurvitch \cite{Gurvitch},
  who has shown that a combination of a large $\lambda$ and significant 
  disorder can combine to produce an enhanced $T^2$ term.   
  
  In the absence of any other likely explanations, 
  we speculate that the wide range of quadratic resistivity may be caused by coupling to a series
  of optical phonon modes, broadly spaced in energy that conspire to produce the 
  observed temperature dependence. 
  This situtaion is similar to that observed in MgB$_2$ \cite{Masui}, however these modes would 
  have to give a small contribution to the overall phonon density
  of states, so as to be overlooked by bulk thermodynamic measurements. With the detailed 
  information about the FS geometry presented
  in this study, a thorough theoretical treatment of electron-phonon coupling in
  Ag$_5$Pb$_2$O$_6$ should be possible.  In this respect, a reliable calculation or an 
  experimental measure of the 
  phonon spectrum of this material through infrared absorption or neutron scattering would be 
  extremely valuable.
  
\begin{acknowledgments}

We would like to thank G.G. Lonzarich and P. Littlewood for useful discussions, and T. Oguchi 
for providing us
with his recent band structure calculations prior to publication.  This research is supported
with funds from EPSRC.  M.S. acknowledges support from an NSERC of Canada postdoctoral fellowship, 
while C.B. acknowledges the support of the Royal Society.  
\end{acknowledgments}


\begin{thebibliography}{25}
\bibitem[{foo()}]{footnote}
\bibinfo{note}{While Li and Rb become superconductors under the application of
  pressure, this is only accomplished through structural changes and an
  $s$-to-$p$ (Li) or $s$-to-$d$ (Rb) transition of the valence electron [7].}
\end{thebibliography}
\end{document}